\documentclass[a4paper, pra, amsmath, showpacs, preprintnumbers,
superscriptaddress, twocolumn, sort&compress, floatfix]{revtex4}
\usepackage[latin1]{inputenc}
\usepackage{hyperref}
\usepackage{graphicx, color}
\usepackage{mathptmx}
\usepackage{textcomp}

\newcommand{\azero}{\ensuremath{a_0}}{}
\newcommand{\lv}[1]{\ensuremath{|#1\rangle}}
\newcommand{\Kinel}{\ensuremath{K_\text{inel}}}
\newcommand{\seeApp}{Appendix} 

\begin{document}

\title{Inducing an Optical Feshbach Resonance via Stimulated Raman
  Coupling}

\author{Gregor Thalhammer}
\author{Matthias Theis}
\author{Klaus Winkler}

\affiliation{Institut f\"ur Experimentalphysik, Universit\"at
  Innsbruck, Technikerstra{\ss}e 25, 6020 Innsbruck, Austria}

\author{Rudolf Grimm}
\affiliation{Institut f\"ur Experimentalphysik, Universit\"at
  Innsbruck, Technikerstra{\ss}e 25, 6020 Innsbruck, Austria}
\affiliation{Institut f\"ur Quantenoptik und Quanteninformation,
  \"Osterreichische Akademie der Wissenschaften, 6020 Innsbruck, Austria}

\author{Johannes Hecker Denschlag}
\affiliation{Institut f\"ur Experimentalphysik, Universit\"at
  Innsbruck, Technikerstra{\ss}e 25, 6020 Innsbruck, Austria}

\date{\today}


\begin{abstract}
  We demonstrate a novel method of inducing an optical Feshbach
  resonance based on a coherent free-bound stimulated Raman
  transition.  In our experiment atoms in a $^{87}$Rb Bose-Einstein
  condensate are exposed to two phase-locked Raman laser beams which
  couple pairs of colliding atoms to a molecular ground state.  By
  controlling the power and relative detuning of the two laser beams,
  we can change the atomic scattering length considerably. The
  dependence of scattering length on these parameters is studied
  experimentally and modelled theoretically.
\end{abstract}

\maketitle

\section{Introduction}
\label{introduction}

Feshbach resonances have become a central tool in the physics of
ultra-cold quantum gases during the last years because they allow for
a tuning of the interactions between atoms.  Controlling interparticle
interactions is a central key in many fields of modern physics and is
especially relevant for future applications in quantum computation and
exploring novel many-particle quantum effects. Beautiful experiments
using magnetically tunable Feshbach resonances
\cite{Tiesinga1993,Inouye1998} have been performed, ranging from
ultra-high resolution molecular spectroscopy \cite{Chin2000} to the
coherent coupling of atomic and molecular states \cite{Donley} as well
as the creation of bright matter wave solitons \cite{Kaykovich}. It
also led to the production of new atomic \cite{Cornish} and molecular
\cite{Jochim} Bose-Einstein condensates (BEC) and allowed to control
pairing in ultra-cold fermionic gases \cite{fermipairs}.

Recently we demonstrated how atom-atom interactions in a $^\text{87}$Rb BEC
can also be tuned with an optically induced Feshbach resonance
\cite{OptFesh1} (see also \cite{Fatemi}), a scheme which was
originally proposed by Fedichev \textit{et al.}
\cite{Fedichev1996,Bohn}. Optically induced Feshbach resonances offer
advantages over magnetically tuned Feshbach resonances since intensity
and detuning of optical fields can be rapidly changed.  Furthermore
complex spatial intensity distributions can be easily produced and
optical transitions are always available even when no magnetic
Feshbach resonances exist. A disadvantage of optically induced
Feshbach resonance is the inherent loss of atoms due to excitation and
spontaneous decay of the molecular state \cite{OptFesh1}. Typical
lifetimes for excited molecular states are on the order of 10\,ns which
corresponds to a linewidth of $2\pi\times 16\,\text{MHz}$. Evidently,
coupling to molecular states with longer lifetime should improve the
situation. Ground state molecules are stable against radiative decay,
and narrow transition line widths on the order of kHz have been
observed in two-photon Raman photoassociation \cite{Wynar,Rom}. This
raises the question whether it is possible to create optical Feshbach
resonances using stimulated Raman transitions and whether this scheme
might be advantageous compared to the one photon optical Feshbach
resonance.

In this paper we indeed demonstrate that optical Feshbach resonances
can be induced using a coherent two-color Raman transition to a highly
vibrationally excited molecular ground state in a $^{87}$Rb BEC. In
the experiment we show how the scattering length and loss rates can be
tuned as a function of the intensity of the lasers and their detuning
from molecular lines.  We use Bragg spectroscopy \cite{Stenger} as a
fast method to measure the scattering length in our sample
\cite{OptFesh1}.  To fit and analyze our data we use a model by Bohn
and Julienne \cite{BohnJulienne99}. We find that using the Raman
scheme for optically induced Feshbach resonances leads to similar
results in tuning of the scattering length as for the single photon
Feshbach scheme.  The Raman scheme does not lead to an improvement
compared to the one-photon scheme because its atomic loss rate is not
lower for a given change in scattering length. However, using a
stimulated Raman transition does offer experimental advantages. To
tune over the Feshbach resonance, the relative frequency of the two
laser beams only has to be changed typically by several MHz which can
be conveniently done using an acousto-optic modulator.  This allows
for very fast and precise control of the scattering length. On the
other hand, working with a one-photon optical Feshbach resonance in
the low loss regime typically requires large detunings and scan ranges
on the order of GHz. The Raman scheme relaxes the necessity for
absolute frequency control of the lasers which can be tedious to
maintain far away from atomic lines. Since off-resonant light fields
in general lead to dipole forces acting on the atoms, a variation of
the scattering length via optical tuning leads to a variation of the
dipole forces on the atomic sample. This unwanted effect can be made
negligible for the Raman scheme which tunes over resonance within a
small frequency range.

The paper is organized as follows: We start in section
\ref{sec:Raman-scheme} by discussing the Raman scheme with a simple
theoretical model. In section \ref{sec:experimental-setup} we describe
in detail our experimental setup and the measurement method. In
section \ref{sec:results} we discuss the experimental results which
are compared with a theoretical model. The appendix gives details of
the model that is used to describe the data.

\section{Raman scheme for optical Feshbach tuning}
\label{sec:Raman-scheme}

\begin{figure}
  \input{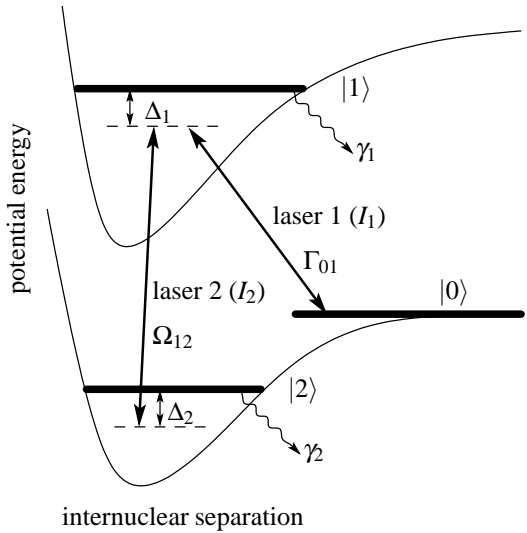}
  \caption{Schematic diagram of the transitions used for optically
    coupling the collisional state $|0\rangle$ to molecular states
    $|1\rangle$ and $|2\rangle$. $|1\rangle$ is electronically
    excited whereas  $|2\rangle$ is in the electronic ground state.
   $\Delta_1$ and $\Delta_2$ are defined to be positive for the
    shown configuration. }
    \label{fig:Raman-scheme}
\end{figure}

Before discussing optical Feshbach tuning based on a two-photon Raman
transition, it is instructive to briefly recall the one-photon scheme
first \cite{OptFesh1,Fedichev1996,Bohn}. This configuration uses a
single laser beam tuned close to a transition from the scattering
state of colliding atoms to a bound level in an excited molecular
potential (states $|0\rangle$ and $|1\rangle$ in
Fig.~\ref{fig:Raman-scheme}). Varying the detuning $\Delta_1$ or the
intensity $I_1$ modifies the coupling and hence the scattering length.
Atomic loss can occur through population of the electronically excited
molecular state which has a decay width of $\gamma_1$.

Introducing a second laser as shown in Fig.~\ref{fig:Raman-scheme}
will now couple the collisional state $|0\rangle$ to a bound level
$|2\rangle$ in the ground state potential. As we will show, this
allows for a tuning of the scattering length similar to the one-photon
scheme.  We now have, however, four parameters which can be used to
influence the scattering length: the intensities $I_1$ and $I_2$ of
lasers 1 and 2 and the detunings $\Delta_1$ and $\Delta_2$ as shown in
Fig.~\ref{fig:Raman-scheme} \footnote{As we observe a significant
  light shift of level $\lv1$, depending on intensity $I_1$ of laser 1
  \cite{OptFesh1}, we measure the detuning $\Delta_1$ from the
  observed position of the one-photon line at a given intensity of
  laser 1. Note that $\Delta_1$ is a one-photon detuning whereas
  $\Delta_2$ is a two-photon detuning.  }.

From \cite[Eqs.~(4.8)--(4.11)]{BohnJulienne99} one can extract
approximate expressions for the inelastic collision rate coefficient
$\Kinel$ and the scattering length $a$ in a Bose-Einstein condensate
\footnote{$\Kinel$ is reduced by a factor of 2 as compared to the case
  of thermal atoms.  This is because in a BEC all atoms share the same
  quantum state.}:
\begin{eqnarray}
  \label{eq:Kinel-3-level}
  \Kinel = \frac{2\pi\hbar}{m} \frac{1}{k_\text{i}}
  \frac{\Gamma_{01}\gamma_1}
  {(\Delta_1^{} - \Omega_{12}^2/\Delta_2^{})^2 + (\gamma_1/2)^2}
  \\
  \label{eq:ascatt-3-level}
  a = a_\text{bg} -
  \frac{1}{2 k_\text{i}} \frac {\Gamma_{01} (\Delta_1^{} - \Omega_{12}^2/\Delta_2^{})}
  {(\Delta_1^{} - \Omega_{12}^2/\Delta_2^{})^2 + (\gamma_1/2)^2}
\end{eqnarray}
Here $\Gamma_{01}$ denotes the on-resonance stimulated transition rate
from $|0\rangle$ to $|1\rangle$ and is proportional to $I_1$.
$\Omega_{12}$ is the Rabi frequency for the coupling of the states
$\lv1$ and $\lv2$ and is proportional to $\sqrt{I_2}$.
$\hbar k_\text{i}$ is the relative momentum of the collision, where
$\hbar$ is Planck's constant divided by $2\pi$.  $a_\text{bg}$ is the
background scattering length and $m$ is the atomic mass.

Equations~\eqref{eq:Kinel-3-level} and \eqref{eq:ascatt-3-level}
neglect spontaneous decay from state~$\lv2$ ($\gamma_2 = 0$) and
assume $\Gamma_{01}\ll\gamma_1$. Setting $\Omega_{12}=0$ yields the
expressions for the one-photon Feshbach resonance as given in
\cite{OptFesh1}.
\begin{figure}
  \centering
  \includegraphics{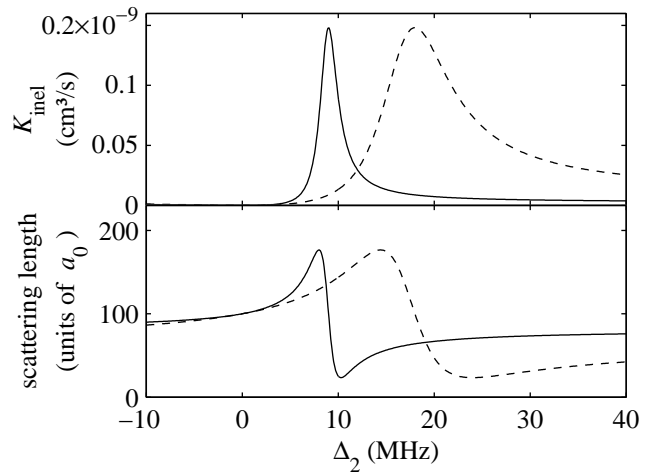}
  \caption{$\Kinel$ and scattering length $a$ according to
    Eqs.~\eqref{eq:Kinel-3-level} and~\eqref{eq:ascatt-3-level},
    plotted for two values of $\Delta_1$. Solid line:
    $\Delta_1/2\pi = 100\,\text{MHz}$, dashed line:
    $\Delta_1/2\pi = 50\,\text{MHz}$. The other parameters are
    $\Gamma_{01}/2\pi = 50\,\text{kHz}$,
    $\Omega_{12}/2\pi = 30\,\text{MHz}$,
    $\gamma_1/2\pi = 25\,\text{MHz}$. The wavenumber
    $k_\text{i} = 2.5\times 10^5\,\text{m}^{-1}$ corresponds to
    the finite size of
     the condensate wavefunction. $a_0$ is the Bohr radius.}
  \label{fig:theory-Kinel-a}
\end{figure}
Equations~\eqref{eq:Kinel-3-level} and \eqref{eq:ascatt-3-level} yield
a Lorentzian and a corresponding dispersive line shape as a function
of $\Delta_1$. In our experiments, however, we hold $\Delta_1$
constant and scan $\Delta_2$.  Figure~\ref{fig:theory-Kinel-a} shows
typical curves for $\Kinel$ and $a$ for two detunings $\Delta_1$. The
curves for $\Kinel$ are slightly asymmetric, but for
$\Delta_1 \gg \Omega_{12}$ they can be well approximated by
Lorentzians. This can be seen by expanding the denominator of
Eq.~\eqref{eq:Kinel-3-level} in terms of $\Delta_2$ at the resonance
position. A light shift displaces the position of the resonance to
${\Omega_{12}}^2/\Delta_1$. It is also interesting to note that the
resonance width decreases with increasing detuning $\Delta_1$ as
$\gamma_1 \left( {\Omega_{12}}/{\Delta_1} \right)^2$.

In a sense the two-photon Raman-Feshbach resonance can be coined
in terms of a one-photon Feshbach scheme. The detuning $\Delta_2$
 effectively replaces the detuning $\Delta_1$ of the one-photon Feshbach
scheme \footnote{There is even a more direct way to understand the
  two-photon Feshbach resonance in terms of a one-photon Feshbach
  resonance.  Laser 1 couples the collision state $\lv0$ to a virtual
  level $\lv{2^\prime}$, which is generated by laser 2 acting on level
  $\lv{2}$. The splitting between $\lv{2^\prime}$ and $\lv{1}$ is given
  by $ \Delta_{2^\prime} = \Delta_1 - \Delta_2$. Its linewidth is
  $\gamma_1 \left( {\Omega_{12}}/{\Delta_{2^\prime}} \right)^2$ and the
  transition rate
  $\Gamma_{02^\prime} = \Gamma_{01} \left(
    {\Omega_{12}}/{\Delta_{2^\prime}} \right)^2$.  }.

Since  Eqs.\,\eqref{eq:Kinel-3-level} and
\eqref{eq:ascatt-3-level} have exactly the same form as for the
one-photon Feshbach resonance, it follows that, given a fixed
free-bound transition rate $\Gamma_{01}$, the maximum tuning range
of the scattering length for the two-photon case cannot be larger
than in a one-photon scheme. Furthermore,  given a fixed
change-in-scattering-length, the loss rate as determined by
$\Kinel$ is not lower for the Raman scheme than for the one-photon
scheme.

\section{Experimental setup and methods}
\label{sec:experimental-setup}

\subsection{Production of BEC}
\label{sec:production-bec}

For the experiments we produce $^{87}$Rb BECs of typically
$1.2\times 10^6$ atoms in the spin state $| F = 1, m_F = -1 \rangle$.
Our setup comprises a magnetic transfer line \cite{Greiner2001} to
transport atoms from a magneto-optic trap (MOT) chamber to a glass
cell where the BEC is produced and all experiments are carried out.
In a first step about $3\times10^{9}$ atoms are loaded within 4\,s
into a MOT directly from the background gas and are then cooled
further to about 50\,\textmu K in a molasses cooling phase. After
optically pumping into the $| F = 1, m_F = -1 \rangle$ state we load
the atom cloud into a magnetic quadrupole trap with a gradient of
130\,G/cm in the (strong) vertical direction.
Within 1.4\,s the atoms are then moved via a magnetic transfer line
\footnote{For our magnetic transport (similar to that described in
  \cite{Greiner2001}) 13 pairs of quadrupole coils are used.  These
  transfer coils each have an inner diameter of 23.6\,mm, an outer
  diameter of 65\,mm, a height of 5.7\,mm and consist of 34 windings.
  They are arranged in two layers above and below the vacuum chamber
  with a separation of 50\,mm.  Peak currents of 75\,A are necessary
  to maintain a vertical gradient of 130\,G/cm during transfer.} %
over a distance of 48\,cm including a 120\textdegree{} corner into a
glass cell which is at a pressure below $10^{-11}\text{ mbar}$.
In this cell we finally load the cloud into a QUIC trap \cite{QUIC}
ending up with typically $4\times 10^8$ atoms at a temperature of
about 250\,\textmu{}K. All three coils of the QUIC trap are operated
at a current of 40\,A, dissipating 350\,W. This results in trap
frequencies of $\omega_\text{radial}/2\pi = 150\,\text{Hz}$ and
$\omega_\text{axial}/2\pi = 15\,\text{Hz}$ at a magnetic bias field of
2 G. To achieve Bose-Einstein condensation we use forced
radio-frequency evaporation for a period of 20\,s. The stop frequency
is chosen so that we end up with condensates with a thermal background
of about 25\% of non-condensed atoms.  At this value we concurrently
get the highest number of atoms in the condensate and good
reproducibility. For our measurements we consider only the condensed
atoms.

\subsection{Raman lasers}
\label{sec:raman-scheme}

To realize the Raman scheme shown in Fig.~\ref{fig:Raman-scheme} we
use the electronically excited molecular state
$|1\rangle = | 0_g^-, \nu = 1, J = 2 \rangle$ located 26.8\,cm$^{-1}$
below the $(S_{1/2} + P_{3/2})$ dissociation asymptote
\cite{Fioretti,OptFesh1}.  About 290\,MHz below the $J = 2$ line,
there is another rotational level with $J = 0$ %
\footnote{Due to different light shifts \cite{OptFesh1} for the $J=0$
  and $J=2$ lines, their splitting is intensity dependent. The value
  of 290\,MHz is valid for an intensity of
  300\,W/cm\textsuperscript{2}.}. %
Although about five times weaker than the $J=2$ line, its effect
cannot be totally neglected in our experiment. We choose level
$|2\rangle$ to be the second to last bound state in the ground state
potential. It has a binding energy of 636\,MHz$\times h $ \cite{Wynar}
where $h$ is Planck's constant.

\begin{figure}
  \centering\input{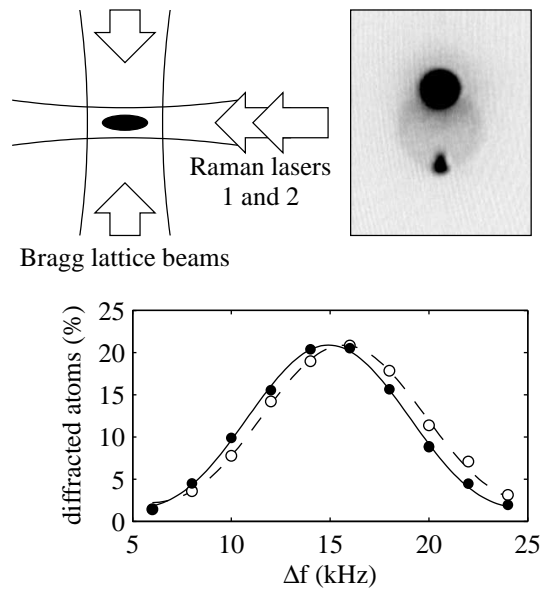}
  \caption{Top left: Experimental arrangement of the laser beams (top
    view). Top right: Absorption image obtained after
    Bragg-diffracting a portion of the atoms to a state with a
    momentum of two photon recoils (lower atom cloud) and subsequent
    time of flight expansion.  Bottom: Bragg resonance curves for two
    different relative detunings of the Raman lasers.  The relative
    shift of 700\,Hz is due to two different scattering lengths which
    are optically induced in the condensates.  The atom numbers are
    the same for both curves.  Shown is the percentage of diffracted
    atoms versus the frequency difference of the Bragg lattice beams.
    For better comparison we have scaled up the right curve by 10\,\%.
  }
 \label{fig:SchemeSetup}
\end{figure}

The Raman laser beams are derived from a Ti:Sapphire laser using an
acousto-optical modulator at a center frequency of about 318 MHz in a
double-pass configuration. This allows precise control of their
relative frequency difference over several tens of MHz.
Both Raman lasers propagate collinearly and are aligned along the weak
axis of the magnetic trap (see Fig.~\ref{fig:SchemeSetup}).  They have
a 1/e$^2$ waist of 76\,\textmu{}m, and their linear polarization is
perpendicular to the magnetic bias field of the trap.

The Ti:Sapphire laser is intensity stabilized and its frequency has a
line width of about 3\,MHz. In order to stabilize its frequency
relative to the photoassociation lines, the laser is offset locked
relative to the $D_2$ line of atomic rubidium with the help of a
scanning optical cavity. This yields an absolute frequency accuracy of
better than 10\,MHz. In all our experiments the Raman laser
intensities were set to $I_1 = 300$ W/cm$^2$ and $I_2 = 60$ W/cm$^2$
at the location of the condensate, if not stated otherwise.

\subsection{Bragg spectroscopy}
\label{sec:bragg-spectroscopy}

To measure optically induced changes in the scattering length $a$, we
use Bragg spectroscopy \cite{OptFesh1,Stenger}.  This method allows
for a fast measurement on time scales below 100\,\textmu{}s which is
vital because of the rapid photoassociation losses we experience in
our experiments. A moving lattice composed of two counter propagating
beams with wavenumber $k$ and a frequency difference $\Delta f$ is
used to diffract some of the condensate atoms to a state with non-zero
momentum. When energy and momentum conservation are fulfilled, the
Bragg lattice resonantly transfers a momentum of two photon recoils
$2\hbar k$ in a first order diffraction process. For the case of a
homogenous condensate of density $n$, the resonance energy for Bragg
diffraction is given by the sum of transferred kinetic energy
$ h \Delta f_0 = (2\hbar k)^2/2m $ and the change in mean-field energy
$ 4\pi \hbar^2 n a /m$ \footnote{This is valid in the limit that only
  a small fraction of the condensate is diffracted.}. This corresponds
to a frequency difference of the Bragg lasers of
\begin{equation}
  \label{eq:braggspectroscopy}
  \Delta f_\text{r} = \Delta f_0 +
  \frac{2 \hbar}{m} n a .
\end{equation}

If the condensate is initially not at rest, the kinetic energy
contribution $\Delta f_0$ to the Bragg resonance frequency (Eq.
\ref{eq:braggspectroscopy}) contains an additional term
$ 2 \hbar k p/m$, where $p$ is the initial atom momentum in the
direction of the Bragg lattice. In our experiments we observe such a
motional shift corresponding to condensate momentum of up to
$p=0.1\,\hbar k$. This momentum can partly be attributed to optical
dipole forces of Raman beams which are slightly non-centered on the
condensate.  Partly it can be attributed to a forced oscillation of
the condensate in the magnetic trap at 150 Hz which coincides with the
trapping frequency.  Since this oscillation is driven by a higher
harmonic of the line frequency (50 Hz), it is in phase with the line
frequency and we are able to stabilize the initial condensate momentum
by synchronizing the experiment to the line. A stable initial
condensate momentum can then be determined and canceled out by
measuring $\Delta f_\text{r}$ alternately for Bragg diffraction to the
$+2\hbar k$ and $-2\hbar k$ momentum component. After these measures
we were left with a residual momentum noise level of up to
$p=0.01\,\hbar k$.

In our setup the Bragg lattice beams are oriented along the horizontal
direction perpendicular to the Raman laser beams (see
Fig.~\ref{fig:SchemeSetup}) and have a width of $\approx 0.9$\,mm.  We
extract both beams from a single grating stabilized diode laser and
use two acousto-optical modulators to control the frequency
difference.  The laser is tuned 1.4\,nm below the $^{87}$Rb $D_2$ line
which defines $\Delta f_0$ in Eq. \ref{eq:braggspectroscopy} to be
15.14\,kHz. This frequency is much larger than the typical mean field
contribution, $2 \hbar n a/m$, which in our experiments was below
3\,kHz.

We illuminate the trapped condensate for 100\,\textmu{}s with the
Bragg lattice light.  After 12\,ms of free expansion the diffracted
atoms are spatially separated from the remaining atoms.  Absorption
imaging allows us to determine the diffraction efficiency. By
adjusting the Bragg laser intensity (typically 1\,mW) we keep the
maximum diffraction efficiency between 15\,\% and 20\,\%.
When we scan the frequency difference $\Delta f$ and measure the
fraction of Bragg diffracted atoms we obtain curves as shown in
Fig.~\ref{fig:SchemeSetup} (bottom).
These curves have a width of approx. 9\,kHz as determined by the
100\,\textmu{}s length of our Bragg pulses. The shape of the curves is
given by the Fourier transform of our square light pulses which we use
to fit the data to obtain the resonance position $\Delta f_\text{r}$
\cite{OptFesh1}. The shift between the two Bragg spectroscopy curves
in Fig.~\ref{fig:SchemeSetup} (bottom) is optically induced by shining
in the Raman lasers at the same time as the Bragg lattice.  For both
curves the atom numbers are the same and $\Delta_1 = 60$ MHz. Only the
Raman detuning $\Delta_2$ differs by 26 MHz.  According to
Eq.\,(\ref{eq:braggspectroscopy}) this observed shift in Bragg
resonance frequency is then due to a change in scattering length,
induced by tuning $\Delta_2$. This demonstrates that we can tune the
scattering length $a$ with a Raman Feshbach resonance.

\subsection{Determination of Scattering Length}
\label{sec:determ-scatt-length}

We use Eq.~\eqref{eq:braggspectroscopy} to determine the scattering
length $a$ from the measurements of the Bragg resonance frequency
$\Delta f_\text{r}$. Equation~\eqref{eq:braggspectroscopy}, however,
is derived for the case of a homogenous condensate. Our trapped
condensate, in contrast, which is subject to photoassociation losses
exhibits a time and position dependent density $n$. This can be taken
into account by replacing the density $n$ in
Eq.~\eqref{eq:braggspectroscopy} by an appropriate effective value
$\bar{n}$. %

A simple approach to estimate $\bar{n}$ is to calculate the spatial
and time average of the condensate density $n$ over the duration of
the Raman pulse length $T$. For this we use the rate equation for the
local density $\dot{n}=-2\Kinel \ n^2$ for two-atom losses. The
inelastic collision rate coefficient $\Kinel$ governing this process
is obtained from measuring the atom number at the beginning and the
end of the light pulse. This procedure already yields good results
which differ less than 10\,\% from an improved approach which we use
for our data analysis and which is explained in the following.

The improved approach consists of a full numerical simulation which
describes Bragg diffraction in a dynamically and spatially resolved
way. We divide the condensate into density classes and and treat their
time dependence individually. The Bragg diffraction process is
identified as a Rabi oscillation between a coherent two level system,
i.e., the BEC component at rest and the Bragg diffracted component.
The changing density of the condensate due to loss is reflected in a
time dependent resonance frequency (see
Eq.~\eqref{eq:braggspectroscopy}).  As a result of these calculations
we obtain for each density class a Bragg resonance curve similar to
the experimental ones shown in Fig.~\ref{fig:SchemeSetup}. Averaging
over these resonance curves and determining the center position yields
the simulated value for the Bragg resonance $\Delta f_\text{r}$. Using
$\Delta f_\text{r} = \Delta f_0 + 2 \hbar \bar{n} a /m $ we can then
determine the effective density $\bar{n}$.

\section{Results}
\label{sec:results}
\subsection{Raman Scans}
\label{sec:scann-raman-detun}

\begin{figure}
  \includegraphics{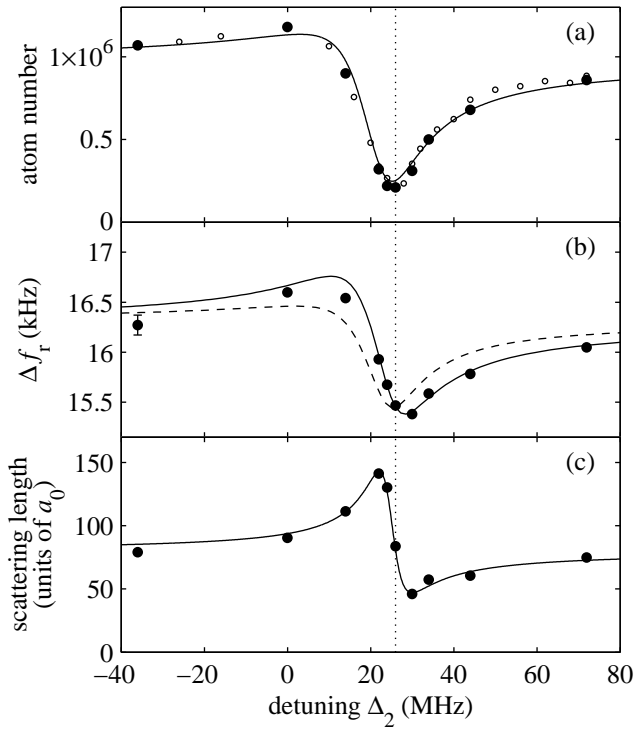}
  \caption{Optical Feshbach resonance using a Raman scheme.  (a) shows
    the measured atom number after the Raman pulse, (b) the measured
    Bragg resonance frequency and (c) the scattering length, as
    determined from (a) and (b).  In (a) the filled circles correspond
    to measurements where Bragg spectroscopy was used to determine the
    scattering length, while the small open circles stem from
    additional loss measurements without Bragg spectroscopy.
    From our measurements we estimate the uncertainty of the
    Bragg resonance frequency to be smaller than $\pm100\,\text{Hz}$,
    as indicated by the error bar in (b).
    The solid lines in (a), (b) and (c) are from a model calculation
    (see Appendix and text).  The dashed line in (b) shows the
    expected signal if there was only loss in atom number but no
    change in scattering length (see also discussion in text).  The
    vertical line indicates the location of maximal loss in (a) and
    helps to compare the relative positions of the three curves.}
\label{fig:ScatteringLength}
\end{figure}

Figure~\ref{fig:ScatteringLength} presents measurements where the
detuning $\Delta_1$ of laser 1 from the excited molecular state is set
to $\Delta_1/2\pi= 60$\,MHz. The intensities of the Raman lasers~1
and~2 are $300$\,W/cm$^2$ and $60\,\text{W/cm}^2$, respectively.
Fig.~\ref{fig:ScatteringLength}(a) shows the atom number after
illuminating a condensate of initially $1.4 \times 10^6$ atoms for
100\,\textmu{}s with the Raman lasers.  Scanning the Raman detuning
$\Delta_2$ we find a strong loss of atoms on resonance.  As already
expected from Eq.~\eqref{eq:Kinel-3-level} the line shape is slightly
asymmetric. Figure~\ref{fig:ScatteringLength}(b) shows the resonance
frequency $ \Delta f_\text{r} $ as measured by Bragg spectroscopy.
When we analyze the data in Fig.~\ref{fig:ScatteringLength}(a) and (b)
with the improved procedure described in
section~\ref{sec:determ-scatt-length} we obtain values for the
scattering length which are shown in
Fig.~\ref{fig:ScatteringLength}(c). The scattering length $a$ shows a
dispersive variation between 50\,\azero{} and 140\,\azero{} as we scan
over the resonance. The dispersive scattering length curve is offset
by about 20 $a_0$ from the background scattering length
$a_{\text{bg}}=100\,\azero$ for $^{87}$Rb in the
$| F = 1, m_F = -1 \rangle$ state
\cite{privJulienneTiesinga,privTiemann,julienne}. This is due to the
one-photon Feshbach tuning of laser 1, in agreement with our previous
measurements \cite{OptFesh1}.

We find that Eqs.~\eqref{eq:Kinel-3-level} and
\eqref{eq:ascatt-3-level} are not sufficient to describe these data
properly, mainly because they neglect the decay rate $\gamma_2$. A
more complete model (see \seeApp), also taking into account both the
$J=0$ and $J=2$ rotational levels, was used for creating fit curves
\footnote{The resulting fit parameters are similar to those given in
  the Appendix.}, depicted as solid lines in
Fig.~\ref{fig:ScatteringLength}. The fact that the data for atomic
loss as well as for the scattering length $a$ are both well described
by the theoretical curves is an intrinsic consistency check for our
model and our data analysis.

The shape of the signal $\Delta f_\text{r}$ in
Fig.~\ref{fig:ScatteringLength}(b) is a combination of the effects of
the varying scattering length $a$ and the varying atom number (see
Eq.~\eqref{eq:braggspectroscopy}). This is illustrated by the dashed
and continuous lines in Fig.~\ref{fig:ScatteringLength}(b): The dashed
line shows the expected signal if only the variations in atom number
would occur and the scattering length stayed constant \footnote{To
  account for the one-photon Feshbach tuning of laser 1, a value for
  the background scattering length $a_\text{bg} = 80\,a_0$ was used for the
  calculation.}.  The continuous line takes both the variations in
atom number and in scattering length into account.  The deviation of
the measured data points from the dashed line is due to an optical
induced change of the scattering length.

\subsection{Dependence on detuning}
\label{sec:dependence-detuning}
\begin{figure}
  \includegraphics{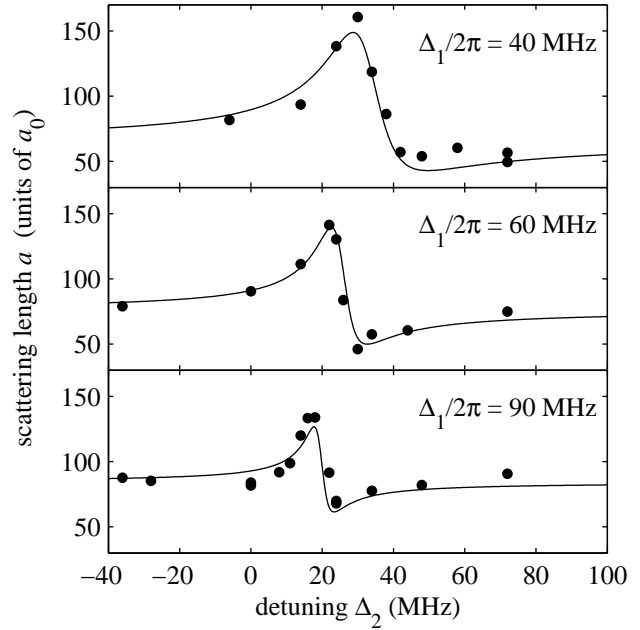}
  \caption{Variation of the scattering length with Raman detuning for
    three various detunings $\Delta_1$ from the excited molecular
    state. The solid line is a calculation (see \seeApp) which uses a
    single set of parameters for all curves.}
  \label{fig:ScatteringLength-vs-detuning}
\end{figure}
\begin{figure}
  \includegraphics{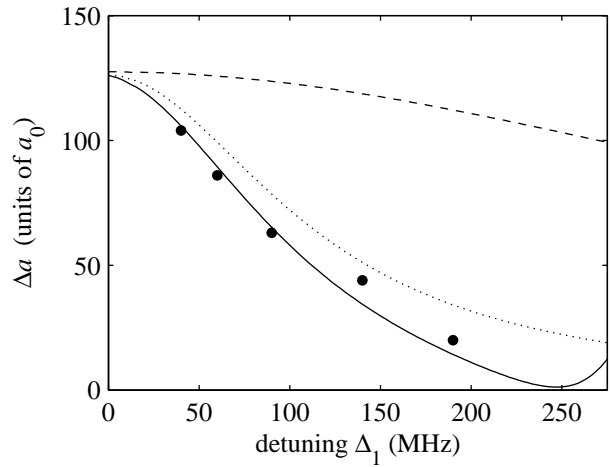}
  \caption{Maximum variation in scattering length
    $\Delta a=a_\text{max}-a_\text{min}$ versus one-photon detuning
    $\Delta_1$.  Solid line: full model calculation (see \seeApp).
    Dotted line: 3-level model (see Fig.~\ref{fig:Raman-scheme}), with
    $\gamma_2/2\pi = 2\,\text{MHz}$.  Dashed line: 3-level model, with
    $\gamma_2/2\pi= 100\,\text{kHz}$.  }
  \label{fig:Delta-a-vs-detuning}
\end{figure}

We now investigate how detuning $\Delta_1$ affects the scattering
length $a$. Figure~\ref{fig:ScatteringLength-vs-detuning} shows a set
of three curves showing the scattering length for detunings
$\Delta_1/2\pi = 40$, $60$ and $90$\,MHz.

The measurements clearly show that the position and width of the
resonances depend on $\Delta_1$. The change in position can be mainly
explained as light shifts of levels $\lv1$ and $\lv2$ due to laser~2.
The decrease of the resonance width with increasing detuning
$\Delta_1$ follows directly our discussion in section
\ref{sec:Raman-scheme}.  The solid lines are model calculations as
described in detail in the \seeApp. They are derived from a
simultaneous fit to the data shown in
Fig.~\ref{fig:ScatteringLength-vs-detuning} and a large number of atom
loss measurements with different detunings (not shown). The set of fit
parameters is listed in Appendix \ref{sec:6-level-model}. We also use
this same set of parameters for the theoretical curves in
Figs.~\ref{fig:Delta-a-vs-detuning} and
\ref{fig:Delta-a-vs-intensity}.

It is interesting to note from
Fig.~\ref{fig:ScatteringLength-vs-detuning} that the amplitude of the
dispersive scattering length signal decreases as $\Delta_1$ becomes
larger. This is not to be expected from the simple model
Eqs.~\eqref{eq:Kinel-3-level} and~\eqref{eq:ascatt-3-level}.
To investigate this effect we have performed scans for atom loss and
scattering length for several detunings $\Delta_1$.
Figure~\ref{fig:Delta-a-vs-detuning} shows the maximum variation in
scattering length, $\Delta a=a_\text{max}-a_\text{min}$, obtained for
detunings $\Delta_1$ ranging from 40\,MHz to 200\,MHz. Here,
$a_\text{max}$ and $a_\text{min}$ are the maximal and minimal
scattering length values for corresponding scan curves. Typical scan
curves are shown in Fig.~\ref{fig:ScatteringLength-vs-detuning}. Each
data point in Fig.~\ref{fig:Delta-a-vs-detuning} was derived from a
complete scan and corresponds to one day of data collection.

An analysis of our data using our theoretical model indicates that the
decrease of $\Delta a$ as a function of $\Delta_1$ is a consequence of
two effects.

(i) To properly model these measurements we have to assign to the
molecular state $\lv2$ in the ground state potential a non-negligible
decay width $\gamma_2/2\pi \approx 2\,\text{MHz}$.  For comparison,
two calculations of a 3-level model are plotted in
Fig.~\ref{fig:Delta-a-vs-detuning}. For small
$\gamma_2/2\pi = 100\,\text{kHz}$ (dashed line) $\Delta a$ decreases
only weakly.  For $\gamma_2/2\pi = 2\,\text{MHz}$ (dotted line) the
theory fits the data much better.
Such a large decay rate of a ground state level is surprising. It
seems too large to be explained purely by collisions. We find that the
decay rate increases with the light intensity. At low light powers of
a few W/cm$^2$ we have observed very narrow linewidths $\gamma_2/2\pi$
on the order of a few kHz, similar to the values reported by
\cite{Wynar,Rom}. The broadening of the molecular ground level could
be due to coupling to excited molecular levels.  We can exclude,
however, from our experimental data that these levels are located
within our experimental scanning range between the states \lv{1} and
\lv{3}. This would lead to additional resonance features in the
scattering length, absorption and light shifts, which are inconsistent
with our data. In contrast, our data indicate a relatively constant
background loss rate of the ground level over the experimental scan
range. This allows us to analyze the data successfully with our simple
few-level model.  Besides coupling to excited molecular states, we
suspect that coupling to the \textit{d}-wave shape resonance of the
scattering channel also gives rise to a sizeable contribution to the
molecular decay rate. Because the \textit{d}-wave shape resonance is
located very close (a few MHz) to threshold, it is resonantly coupled
to the molecular ground state level via the Raman transition. To
include the shape resonance is beyond the reach of our simple model
and has to be investigated later.

(ii) The second reason for the decrease in $\Delta a$ is a quantum
interference effect involving both the $J=2$ and $J=0$ rotational
levels as predicted by our model. At a detuning of
$\Delta_1/2\pi \approx 250\,\text{MHz}$ the interference effect leads
to a complete disappearence of the optical Feshbach resonance. We
observe this in a corresponding disappearence of the atom loss feature
in our measurements (not shown). The interference effect alone, i.e.,
without a 2 MHz linewidth, is not sufficient to explain the
experimental data in Fig.~\ref{fig:Delta-a-vs-detuning}.

\subsection{Dependence on intensity}
\label{sec:depend-intens}

From the simple model Eq.~\eqref{eq:ascatt-3-level} it is clear that
the maximum variation in scattering length $\Delta a$ is proportional
to $\Gamma_{01}$ and consequently scales linearly with the intensity
$I_1$ of laser 1. We have verified this dependence recently
\cite{OptFesh1} for the case of a one-photon optical Feshbach
resonance.

In contrast, the dependence of $\Delta a$ on intensity $I_2$ of
laser~2 is not so trivial. According to the simple model
Eqs.~\eqref{eq:Kinel-3-level} and~\eqref{eq:ascatt-3-level} which
neglect the decay rate $\gamma_2$, the maximum change $\Delta a$ is
independent of $I_2$. It is also clear, that for $I_2=0$ we have
$\Delta a = 0$ since there is no dependence of scattering length on
$\Delta_2$ at all. This unphysical discontinuous behavior can be
resolved, if we introduce a finite decay rate $\gamma_2>0$. We then
find that for increasing intensity $I_2$, $\Delta a$ rises
continuously from zero to a value where it saturates. We observe this
general behavior in our measurements presented in
Fig.~\ref{fig:Delta-a-vs-intensity}. Our full model, as described in
the Appendix, describes the measured data well if we set the decay
rate to $\gamma_2/2\pi=2\,$MHz (solid line). In contrast, the dashed
line in Fig.~\ref{fig:Delta-a-vs-intensity} shows the calculation for
the same model where $\gamma_2$ is set to $\gamma_2/2\pi=100\,$kHz.
Saturation then occurs at a much lower intensity $I_2$ than for
$\gamma_2/2\pi=2\,$MHz.

\begin{figure}
  \includegraphics{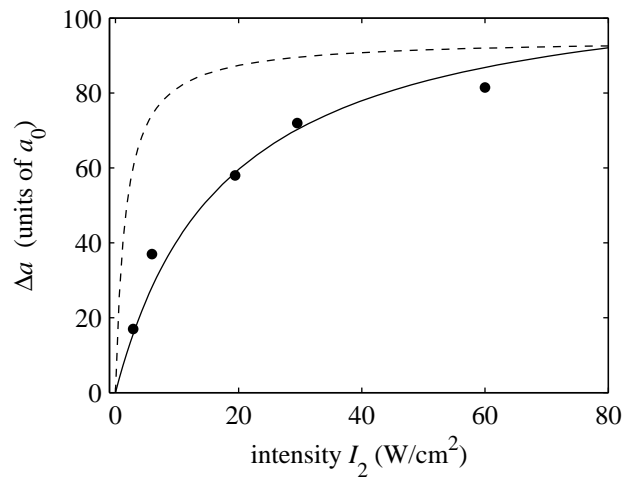}
  \caption{Maximum variation in scattering length
    $\Delta a=a_\text{max}-a_\text{min}$ versus $I_2$. For this data
    set $I_1$ = 300\,W/cm$^2$ and $\Delta_1 =$ 60\,MHz.  The solid
    line is a full model calculation (see \seeApp).  The dashed line
    stems from the same model, but with $\gamma_2/2\pi$ set to
    100\,kHz and is scaled by a factor of 0.84 for better comparison.
  }
  \label{fig:Delta-a-vs-intensity}
\end{figure}

\section{Conclusion}
\label{sec:conclusion}

Our experiments demonstrate the use of an optical Feshbach resonance
for tuning of the scattering length via stimulated Raman coupling to a
bound molecular state. Our results show that there is no advantage
over a one-photon scheme when comparing tuning range and loss rate.
However, for certain applications a Raman scheme is experimentally
more favorable since it demands a lower tuning range of the lasers.
Our presented theoretical model is in good agreement with our data and
might be helpful when tailoring experimental parameters for a specific
application. Furthermore it gives insight into the process of creating
stable ultracold molecules via two-photon photoassociation.

\section{Acknowledgments}
\label{sec:acknowledgements} We appreciate the help of George Ruff and
Michael Hellwig at an early stage of the experiment. We thank Paul
Julienne, Eite Tiesinga, John Bohn, Olivier Dulieu, Peter Fedichev,
Andrea Micheli and Helmut Ritsch for valuable discussions. This work
was supported by the Austrian Science Fund (FWF) within SFB 15
(project part 17) and the European Union in the frame of the Cold
Molecules TMR Network under contract No.\ HPRN-CT-2002-00290.

\appendix
\section{Theoretical model and fit parameters}
\label{sec:theoretical-model}

We use a theoretical model by Bohn and Julienne \cite{BohnJulienne99}
to fit the data in
Figs.~\ref{fig:ScatteringLength}--\ref{fig:Delta-a-vs-intensity}.  In
the following we give a short summary of this model and present the
procedure to calculate the scattering matrix $S$, the loss coefficient
$\Kinel$ and the scattering length $a$. The model has the advantage
that it is concise and intuitive and it allows treatment of multilevel
systems with several couplings between the levels. The numerical
calculations involve simple manipulations of small matrices.

In Fig.~\ref{fig:scheme_four_level} the level scheme for our two
models involving 4 and 6 levels are shown.  We first restrict our
description to the 4-level model as shown in the right part of
Fig.~\ref{fig:scheme_four_level}. In this way our description stays
compact and matrices are kept small. The extension to 6 or more levels
follows the same scheme.

\subsection{The 4-level model}
 \label{subsec:fourlevelmodel}
\begin{figure}
  \centering\input{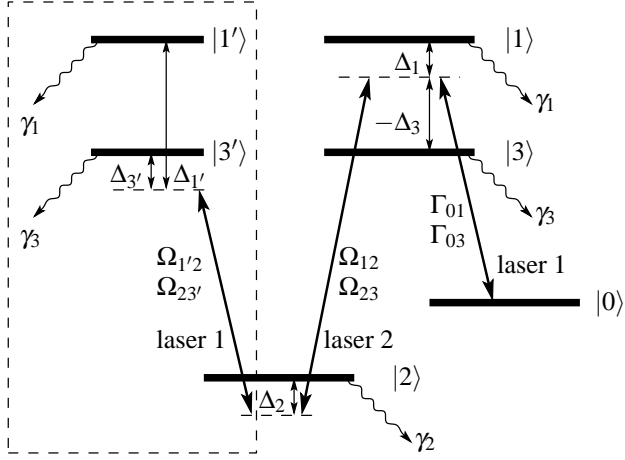}
  \caption{Extended level scheme (compare to
    Fig.\ref{fig:Raman-scheme}) for the 4-level model (right hand
    side) and its extension with 6 levels.  State $\lv{3}$ corresponds
    to the $J=0$ level and lies 290 MHz below the $J=2$ level
    $\lv{1}$. The 4-level model is based on levels $\lv{0}$, $\lv{1}$,
    $\lv{2}$ and $\lv{3}$.  The auxiliary levels $\lv{1^\prime}$ and
    $\lv{3^\prime}$ are introduced in the extended model to describe
    the coupling between $\lv{2}$ and $\lv{1}$, $\lv{3}$,
    respectively, due to laser 1 (see text). }
  \label{fig:scheme_four_level}
\end{figure}

Compared to Fig.~\ref{fig:Raman-scheme} an additional excited level
$\lv3$ is added.  This level corresponds to the rotational level $J=0$
and lies $290\,\text{MHz}$ below the $J=2$ rotational level $\lv1$
\cite{OptFesh1}. We work in the dressed atom picture and every level
$| i \rangle$ is attributed a detuning $\Delta_i$ (see
Fig.~\ref{fig:scheme_four_level}).  $\Delta_0$ is arbitrarily set to
0.  The transition strengths from the continuum $\lv0$ to levels
$\lv1$ and $\lv3$ are described by stimulated rates $\Gamma_{01}$ and
$\Gamma_{03}$ which are proportional to the intensity $I_1$ of laser
1.  The transitions between the bound levels $\lv2$ and $\lv1$, $\lv3$
are characterized by the Rabi frequencies $\Omega_{12}$,
$\Omega_{23}$, respectively, which are proportional to $\sqrt{I_2}$.
Spontaneous decay from the bound levels leading to atomic losses is
formally taken into account by introducing artificial levels
$\lv{\text{a}_i}$ for each level $\lv{i}$ to which a transition at
rate $\gamma_i$ takes place (not shown in
Fig.~\ref{fig:scheme_four_level}).
All these couplings between different levels are summarized in the
symmetric reaction matrix $K$. We arrange the level names in the order
(0, a$_1$, a$_2$, a$_3$, 1, 2, 3) and use them as row and column
indices. The nonzero matrix elements of the $K$ matrix then read
$K_{01} = \sqrt{\Gamma_{01}/2}$, $K_{03} = \sqrt{\Gamma_{03}/2}$,
$K_{i\text{a}_i} = \sqrt{\gamma_i/2}$, $K_{12} = \Omega_{12}$ and
$K_{23} = \Omega_{23}$.  Levels 0, a$_1$, a$_2$, a$_3$ are referred to
as open channels, levels 1, 2 and 3 as closed channels. The reaction
matrix $K$ is partitioned into open and closed channel blocks,
\[
K =
\begin{pmatrix}
  \mathbf{0} &K^\text{oc}\\
  K^\text{co}&  K^{\text{cc}} \\
\end{pmatrix} .
\]
$ K^\text{oc} $ reads in our case
\[
K^\text{oc} =
\begin{pmatrix}
   \sqrt{\Gamma_{01}/2} & 0 & \sqrt{\Gamma_{03}/2}\\
    \sqrt{\gamma_1/2} & 0 & 0\\
    0 & \sqrt{\gamma_2/2} & 0 \\
    0 & 0 & \sqrt{\gamma_3/2}
\end{pmatrix}
. \]
 $K^\text{co}$ is the transposed matrix of $K^\text{oc}$, and
\[
K^\text{cc} =
\begin{pmatrix}
  0 &  \Omega_{12} & 0\\
  \Omega_{12} & 0 &  \Omega_{23}\\
  0 &  \Omega_{23} & 0
\end{pmatrix}.
\]
From $K$ the reduced $K$-matrix,
\[K^\text{red} = K^\text{oc} (D - K^\text{cc})^{-1}  K^\text{co}  \]
is calculated, eliminating the closed channels 1--3, where $D$
denotes a diagonal matrix with diagonal elements $(\Delta_1,
\Delta_2, \Delta_3)$. This determines the unitary $4\times4$
scattering matrix $S$
\[S = ({\mathbf{1} + iK^\text{red}})({\mathbf{1} - iK^\text{red}})^{-1}. \]
From the matrix elements $S_{ij}$ of $S$ the trap loss coefficient
$\Kinel$  is
calculated by
\[
\Kinel = \frac{\pi\hbar}{\mu k_\text{i}}\sum_i \left|S_{0 \text{a}_i}\right|^2
=
\frac{\pi\hbar}{\mu k_\text{i}}
\left(
1 - |S_{00}|^2
\right),
\]
where $\mu = m_\text{Rb}/2$ is the reduced Rb mass and
$\hbar k_\text{i}$ the relative momentum of the colliding atoms. The
scattering length is obtained from $S_{00}$ via
\[
a = a_\text{bg} - \frac{1}{2k_\text{i}} \frac{\Im( S_{00})}{\Re
(S_{00})},
\]
where $\Re( S_{00})$ and $\Im( S_{00})$ denote the real and imaginary
parts of $S_{00}$, respectively.

In the limit of small relative momentum $\hbar k_\text{i}$ and small
coupling strengths $\Gamma_{0i} \ll \gamma_i$, $\Kinel$ and the light
induced change of scattering length $a - a_\text{bg}$ are independent
of $k_\text{i}$ because the $\Gamma_{0i}$ are proportional to
$k_\text{i}$ (Wigner threshold regime) \cite{BohnJulienne99}.

\subsection{Extension of the 4-level model}
\label{sec:6-level-model}

The 4-level model neglects that laser 1 (of which the intensity is
typically five times greater than that of laser 2) also couples the
levels $\lv2$--$\lv1$ and $\lv2$--$\lv3$. However, this coupling
should be taken into account since laser~1 is not far detuned from
these transitions (see Fig.~\ref{fig:scheme_four_level}) due to the
small binding energy of state $\lv3$ ($636\,\text{MHz}\times h$) which
is comparable to typical detunings $\Delta_1$. It mainly leads to
broadening and light shifting of level $\lv2$.  The additional
coupling can approximately be taken care of by adding another two
auxiliary levels $\lv{1^\prime}$ and $\lv{3^\prime}$ with detunings
$\Delta_{1^\prime} = \Delta_{1} + \Delta_{2} + 2\pi\times
636\,\text{MHz}$
and
$\Delta_{3^\prime} = \Delta_{3} + \Delta_{2} + 2\pi\times
636\,\text{MHz}$
as shown in Fig.~\ref{fig:scheme_four_level}. The coupling strengths
$\Omega_{1^\prime2} $ and $\Omega_{23^\prime} $ are fixed by
$\Omega_{1^\prime2} = \Omega_{12} \sqrt{I_1/ I_2} $ and
$\Omega_{23^\prime} = \Omega_{23} \sqrt{I_1/ I_2} $. Compared to the
4-level model no new fit parameters are introduced. We can calculate
$\Kinel$ and the scattering length $a$ following the same recipe as
for the 4-level model, only with larger matrices.
Fitting the data in
Figs.~\ref{fig:ScatteringLength}--\ref{fig:Delta-a-vs-intensity} this
extended model produced much better results than the 4-level model.
For completeness we give here the fit parameters which were used in
the calculations in
Figs.~\ref{fig:ScatteringLength-vs-detuning}--\ref{fig:Delta-a-vs-intensity}
($I_1 = 300\,\text{W}/\text{cm}^2$ and
$I_2 = 60\,\text{W}/\text{cm}^2$):
$\Gamma_{01}/2\pi = 42\,\text{kHz}$,
$\Gamma_{03}/2\pi = 8\,\text{kHz}$,
$\Omega_{12}/2\pi = 32\,\text{MHz}$,
$\Omega_{23}/2\pi = 12\,\text{MHz}$,
$ \gamma_1/2\pi = 25\,\text{MHz}$,
$\gamma_3/2\pi = 22\,\text{MHz}$,
$\gamma_2/2\pi = 2\,\text{MHz}$.  We used
$k_\text{i} = 2.5 \times 10^{-5} m^{-1}$.
Due to the limitations of our model, these fit parameters should not
be mistaken as the true values of the corresponding physical
quantities.


\begin{thebibliography}{99}
\bibitem{Tiesinga1993}
  E. Tiesinga, B.J. Verhaar, and H.T.C. Stoof, Phys. Rev. A
  \textbf{47}, 4114 (1993).

\bibitem{Inouye1998}
  S. Inouye \textit{et al.}, Nature \textbf{392}, 151 (1998);
  Ph. Courteille, R.S. Freeland, D.J. Heinzen, F.A.  van Abeelen and
  B.J. Verhaar, Phys. Rev. Lett. \textbf{81}, 69 (1998);
  J.L. Roberts \textit{et al.}, Phys. Rev. Lett. \textbf{81}, 5109 (1998).

\bibitem{Chin2000}
  C. Chin, V. Vuletic, A.J. Kerman, and S. Chu, Phys. Rev. Lett.
  \textbf{85}, 2717 (2000);
  A. Marte \textit{et al.},
  Phys. Rev. Lett. \textbf{89}, 283202 (2002).


\bibitem{Donley} E.A. Donley, N.R. Claussen, S.T. Thompson and C.E.
  Wieman
  Nature \textbf{417}, 529 (2002).

\bibitem{Kaykovich}
  L. Khaykovich \textit{et al.},
  Science \textbf{296}, 1290 (2002);%
  K.E. Strecker, G.B. Partridge, A.G. Truscott, and R.G. Hulet,
  Nature \textbf{417}, 150 (2002).

\bibitem{Cornish}
  S.L. Cornish, N.R. Claussen, J.L. Roberts, E.A. Cornell, C.E.  Wieman, Phys.
  Rev. Lett. \textbf{85}, 1795 (2000);
  T. Weber, 
  J. Herbig, M. Mark, H.-C. Nägerl, and R. Grimm,
  Science \textbf{299}, 232 (2003), published online 5 December 2002,
  10.1126/science.1079699.


\bibitem{Jochim} S. Jochim \textit{et al.}, Science \textbf{302}, 2101
  (2003); %
  M.  Greiner, C.A. Regal, and D.S. Jin, Nature \textbf{426},
  537 (2003); %
  M. Zwierlein \textit{et al.}, Phys. Rev. Lett. \textbf{91},
  250401 (2003); %
  T. Bourdel \textit{et al.}, Phys. Rev. Lett. \textbf{93},
  050401 (2004).

\bibitem{fermipairs} C.A. Regal, M. Greiner, D.S. Jin, Phys. Rev.
  Lett. \textbf{92}, 040403 (2004); %
  M.W. Zwierlein {\it et al.}, Phys.  Rev. Lett. \textbf{92}, 120403
  (2004); %
  C. Chin {\it et al.}, Science \textbf{305}, 1128 (2004), published
  online 22 July 2004, 10.1126/science.1100818.

\bibitem{OptFesh1}
  M. Theis {\it et al.},
  Phys. Rev. Lett. \textbf{93}, 123001 (2004).

\bibitem{Fatemi} F.K. Fatemi, K.M. Jones, and P.D. Lett,
  Phys. Rev. Lett. \textbf{85}, 4462 (2000).

\bibitem{Fedichev1996} P.O. Fedichev, Y. Kagan, G.V. Shlyapnikov, and
  J.T.M. Walraven, Phys. Rev. Lett \textbf{77}, 2913 (1996).

\bibitem{Bohn} J.L. Bohn and P.S. Julienne, Phys. Rev. A \textbf{56}, 1486
  (1997).

\bibitem{Wynar} R. Wynar, R.S. Freeland, D.J. Han, C. Ryu, D.J.
  Heinzen,
  Science \textbf{287}, 1016 (2000).

\bibitem{Rom} T. Rom \textit{et al.},
  Phys. Rev. Lett. \textbf{93}, 073002 (2004).

\bibitem{Stenger} J. Stenger \textit{et al.},
  Phys. Rev. Lett. \textbf{82}, 4569 (1999).

\bibitem{BohnJulienne99} J.L. Bohn and P.S. Julienne, Phys. Rev. A
  \textbf{60}, 414 (1999).

\bibitem{Greiner2001} M. Greiner, I. Bloch, T.W.  H\"ansch, and T.
  Esslinger,
  Phys. Rev. A \textbf{63}, 031401(R) (2001).

\bibitem{QUIC}
  T. Esslinger, I. Bloch, and T.W. H\"ansch, Phys. Rev. A \textbf{58},
  R2664 (1998).


\bibitem{Fioretti} A. Fioretti \textit{et al.},
  Eur. Phys. J. D \textbf{15}, 189 (2001).

\bibitem{privTiemann} Eberhard Tiemann, private communication.

\bibitem{privJulienneTiesinga} Paul Julienne and Eite Tiesinga,
  private communication.

\bibitem{julienne}
  P.S. Julienne,  
  F.H. Mies, E. Tiesinga, and C.J. Williams, Phys. Rev. Lett.
  \textbf{78}, 1880 (1997).
\end{thebibliography}
\end{document}